\documentclass[aps,pra,twocolumn,showpacs]{revtex4-1}

\usepackage{graphicx,color}
\usepackage{subfigure}
\usepackage{amsmath}
\usepackage[percent]{overpic}

\newcommand{\pa}{\partial}
\newcommand{\total}{\mathrm{d}}

\providecommand{\gtsimeq}{\raisebox{-0.6ex}{$\,\stackrel
	{\raisebox{-.2ex}{$\textstyle >$}}{\sim}\,$}}

\providecommand{\ltsimeq}{\raisebox{-0.6ex}{$\,\stackrel
	{\raisebox{-.2ex}{$\textstyle <$}}{\sim}\,$}}

\begin{document}

\title{Equilibration of a finite temperature binary Bose gas formed by population transfer}

\author{R. W. Pattinson}
\email[]{r.w.pattinson@ncl.ac.uk}
\affiliation{Joint Quantum Centre (JQC) Durham--Newcastle, School of Mathematics and Statistics, Newcastle University, Newcastle upon Tyne, NE1 7RU, United Kingdom}
\author{N. P. Proukakis}
\affiliation{Joint Quantum Centre (JQC) Durham--Newcastle, School of Mathematics and Statistics, Newcastle University, Newcastle upon Tyne, NE1 7RU, United Kingdom}
\author{N. G. Parker}
\affiliation{Joint Quantum Centre (JQC) Durham--Newcastle, School of Mathematics and Statistics, Newcastle University, Newcastle upon Tyne, NE1 7RU, United Kingdom}

\date{\today}

\begin{abstract}
We consider an equilibrium single-species homogeneous Bose gas from which a proportion of the atoms are instantaneously and coherently transferred to a second species, thereby forming a binary Bose gas in a non-equilibrium initial state.   We study the ensuing evolution towards a new equilibrium, mapping the dynamics and final equilibrium state out as a function of the population transfer and the interspecies interactions by means of classical field methods. While in certain regimes, the condensate fractions are largely unaffected by the population transfer process, in others, particularly for immiscible interactions, one or both condensate fractions are vastly reduced to a new equilibrium value.

\end{abstract}


\maketitle
    
  \section{Introduction}

Binary mixtures of atomic Bose--Einstein condensates (BECs) have been the subject of intense research since the first experimental realisation with two sympathetically cooled hyperfine states of $^{87}$Rb~\cite{PhysRevLett.78.586}.  The interplay between the two species, driven by the nonlinear atomic interaction, gives insight into atomic physics \cite{Egorov}, supports rich nonlinear phenomena such as phase separation \cite{Durham,PhysRevLett.101.040402,PhysRevLett.81.1539,PhysRevA.82.033609} and solitons \cite{Becker,Hoefer,Hamner}, and provides an attractive setting for exploring nonequilibrium dynamics \cite{PhysRevLett.81.243,PhysRevLett.99.190402}. 

Binary mixtures (which we treat as distinct from spinor gases \cite{spinor_remark}) have been achieved experimentally with two hyperfine states of the same isotope~\cite{PhysRevLett.78.586,PhysRevLett.81.1539,PhysRevLett.85.2413,McGuirk,PhysRevA.63.051602,PhysRevLett.99.190402,PhysRevA.80.023603,PhysRevA.82.033609,PhysRevLett.82.2228}, with different atomic species~\cite{PhysRevLett.89.053202,PhysRevLett.89.190404,PhysRevLett.100.210402,Durham,EPJD.65.1434,PRA.87.050702,PRA.88.023601} and different isotopes of the same species~\cite{PhysRevLett.101.040402}.  Under the former scenario, it is common to first form a Bose-condensed gas with atoms in a single hyperfine state and then to coherently transfer a population of the atoms into a second hyperfine state.  Most commonly this is performed with $^{87}$Rb and its $|F=1,m_F=-1\rangle$ and $|F=2,m_F=1\rangle$ hyperfine states.  The population transfer is induced by the application of one or more electromagnetic coupling pulses which drive Rabi oscillations between the states, and controllable through the length and amplitude of the pulses.  This approach has been employed to study collective modes of the system  \cite{PhysRevLett.81.243,PhysRevLett.81.1539}, phase dynamics \cite{PhysRevLett.81.1543,Anderson}, pattern formation \cite{PhysRevLett.99.190402}, crossover from triangular to square vortex lattices \cite{Schweikhard} and to measure {\it s}-wave scattering lengths \cite{Egorov}.  The same principle is also exploited for outcoupling atoms to an untrapped state for the production of atom laser beams \cite{Mewes}.  

Importantly, the timescale of the atom transfer (set by the Rabi frequency) is typically an order of magnitude shorter than the timescale of the external dynamics \cite{PhysRevLett.81.243}.  In other words, the atom transfer is effectively instantaneous with respect to the external dynamics, such that the binary mixture is formed in a non-equilibrium state.  In the mean-field context of the ordinary GPE, the dynamics and collective modes of the {\em condensate} following the population transfer have been successfully modelled by a system of coupled Gross--Pitaevskii equations (GPEs -- see e.g. Refs \cite{PhysRevLett.81.243, PhysRevLett.99.190402, Haque}). Within this GPE model, the variable $\psi$ refers directly to the condensate. However, a more fundamental question exists over the {\it thermodynamic} evolution from the non-equilibrium state.  One would anticipate that the instantaneous atom transfer could heat the system and reduce the condensate fraction.  This effect is the motivation for the current work.  

In order to theoretically model thermal excitations and hence describe variations in condensate fraction, one must progress beyond the standard mean-field approximation to a model which describes both the condensate and the thermal atoms in the gas.  Various finite-temperature descriptions of Bose-condensed gases exist (see \cite{Proukakis,finite_temp_book,Blakie} for reviews), briefly classified into two categories: in symmetry-breaking approaches, one maintains a single-mode condensate description, supplementing a slightly amended dissipative GPE for the condensate with a (quantum) Boltzmann equation for the thermal atoms \cite{ZNG_book,ZNG_JLTP}; an alternative approach, which is more commonly implemented numerically, is based on the identification that the main dynamics of interest happen in the low-lying modes.  These modes are typically highly-occupied and can thus be treated classically. This leads to a multi-mode description of the system in which the classical field also obeys the GPE, but now encompassing both the condensate and the low-lying excitations in a unified manner ~\cite{Svis1,Svis2,Svis3,Svis4,PRA.54.5037,Svis5,Davis,OE.8.92,PRL.87.210404, PhysRevA.66.013603,Davis2,PhysRevLett.95.263901,Pol_Rev}. In order to make use of such a classical field description, one needs to have populations initially distributed in modes other than the condensate mode (see also variants of this method~\cite{JLTP.124.431,PRA.65.013603,Blakie,Blakie.arx} based on slightly different initial mode seeding). This can be facilitated by a (non-zero) initial condition across all classical modes, which thus enables mode-mixing during the numerical evolution. Subsequently, the ``classical field'' GPE will thermalize to the classical thermal distribution of the system. 

 The power of the classical field methods is evidenced in their success in modelling phenomena and extracting quantities not accessible to the standard GPE, such as thermal equilibration dynamics~\cite{PhysRevA.66.013603,PhysRevLett.95.263901}, condensate fractions \cite{Davis}, critical temperatures \cite{Davis2006}, correlation functions \cite{Wright2011} and spontaneous production of vortex-antivortex pairs in quasi-2D gases \cite{Simula}.   

The seminal numerical demonstration of this method for a single-species Bose gas (in a 3D periodic box) is presented in Ref.~\cite{PhysRevA.66.013603}.  The simulation begins from a strongly nonequilibrium state where the mode occupation numbers $n_k$ are uniformly distributed over wavenumber $k$ (up to the cutoff in momentum introduced by the numerical grid or a more rigorous cutoff defined by a ``projector" \cite{Blakie}), and the phase of each component is randomized.  At early times the system is weakly turbulent.  Self-ordering of the system leads to the growth of occupation numbers at low $k$.  Over time the mode occupation $n_k$ evolves towards an equilibrium corresponding to the `classical' (Rayleigh--Jeans) limit of the Bose-Einstein distribution $n_k = k_B T/(\epsilon_k-\mu)$, where $\epsilon_k$ is the energy of mode $k$, $T$ is temperature and $\mu$ is the chemical potential \cite{Davis,PRL.87.210404}.  Providing the system is in the regime of $(N,T)-$space for Bose-Einstein condensation to occur, then the equilibrium state consists of a quasi-condensate at low $k$, characterized by macroscopic mode populations (within which the $k=0$ mode is associated with the true condensate, $n_0$), and a thermal non-condensate at high $k$, with low mode occupations.   The quasi-condensate has superfluid ordering and exists in a state of superfluid turbulence, featuring a tangle of quantum vortices which slowly relaxes over time.  The thermodynamic equilibrium state, and hence its characteristic condensate fraction and temperature, is uniquely determined by the initial atom number and energy \cite{PhysRevLett.95.263901}. 

The miscibility/immiscibility of binary Bose gases is a determining factor in their static and dynamic properties.    The species are immiscible when the {\it interspecies} interactions, characterized by the coefficient $g_{12}$, are more repulsive than the {\it intraspecies} interactions, characterized by $g_{11}$ and $g_{22}$.  The immiscibility criteria can be written as~\cite{PhysRevLett.77.3276,Pethick2002},
\begin{equation}
g_{12}^{2}>g_{11}g_{22}. 
\label{eqn:crit} 
\end{equation}
In this regime, it is favourable for the two species to separate spatially, supporting phase-separated equilibrium density profiles in, for example, side-by-side and ball-in-shell formations \cite{Durham,PhysRevLett.101.040402,PhysRevLett.81.1539,PhysRevA.82.033609,PhysRevLett.77.3276,Pu,Trippenbach,Gautum,Jezek,Gordon,Balaz,Pattinson2013,Roy2014}.  From nonequilibrium conditions, complicated dynamical phenomena can also arise driven by the interplay between the two species, such as superfluid ring excitations \cite{PhysRevLett.99.190402}, transient structures during growth \cite{Pattinson2014} and the formation of topological defects in the form of dark-bright solitons \cite{Lui}.  The evolution of binary Bose gases from highly nonequilibrium initial conditions has also been studied previously in the miscible regime \cite{Berloff_2006,Salman20091482}.  

In this work we will study the nonequilibrium dynamics following the separation of an equilibrated, homogeneous Bose-condensed gas into two species, performed via classical field simulations.  In particular we map out how the final equilibrated state of each species, characterized by their condensate fractions, varies with the amount of population transfer and the interspecies interactions.   In Sec.~\ref{sec:theory} we introduce the classical field method of the binary Bose gas, and discuss the generation of the equilibrated single-species gas and the subsequent component separation.  Section~\ref{sec:results} presents our findings, first considering equally-populated components before generalizing to arbitrary populations.  We give our closing remarks in Sec.~\ref{sec:conc}.

  \section{Theoretical Framework}
\label{sec:theory}
\subsection{Classical Field Description}

We consider two weakly-interacting Bose-condensed gases, denoted species 1 and 2.  Assuming that each mode of each species is highly populated then each species can be parametrized by a classical field $\psi_{i}({\bf r},t)$ ($i=1,2$).  Each field provides the density distribution of particles as $\rho_i({\bf r},t)=|\psi_{i}({\bf r},t)|^2$.

 The dynamics of the classical fields are described by the coupled GPEs
  \begin{eqnarray}
i \hbar \frac{\pa\psi_{1}}{\pa t}&=&\left[-\frac{\hbar^2}{2m_1}\nabla^{2}+g_{11}\left|\psi_{1}\right|^{2}+g_{12}\left|\psi_{2}\right|^{2}\right]\psi_{1},\label{cgpe1}\\
  i \hbar \frac{\pa\psi_{2}}{\pa t}&=&\left[-\frac{\hbar^2}{2m_2}\nabla^{2}+g_{22}\left|\psi_{2}\right|^{2}+g_{12}\left|\psi_{1}\right|^{2}\right]\psi_{2}. \label{cgpe2}
  \end{eqnarray}
Here $g_{ii}=4 \pi \hbar^2 a_{ii}/m_i$ parametrizes the strength of the {\it intra}species interactions within species $i$, where $a_{i}$ is the corresponding {\it s}-wave scattering length and $m_i$ is the atomic mass of species $i$;  the {\it inter}species interactions are parametrized by $g_{12}=2 \pi \hbar^2 a_{12}/m_{12}$, where $a_{12}$ is the {\it s}-wave scattering length between the two species and $m_{12}=m_1 m_2/(m_1+m_2)$ is the reduced mass.   Since we are considering the physical situation where the two species are formed as two hyperfine states of the same atom, we take $m_1=m_2=m$.  

We work with the ``natural units'' of species 1: density is expressed in terms of the bulk density of species 1, $\rho_1$, length in terms of the healing length of species 1, $\xi_1=\hbar/\sqrt{2 m g_{11} \rho_1}$, and time in terms of the quantity $\hbar/g_{11} \rho_1$.  Then the coupled GPEs are recast in the dimensionless form
  \begin{equation}
    \begin{split}
      &i\frac{\pa\psi_{1}}{\pa t}=\left[-\nabla^{2}+\left|\psi_{1}\right|^{2}+\frac{g_{12}}{g_{11}}\left|\psi_{2}\right|^{2}\right]\psi_{1},\\
      &i\frac{\pa\psi_{2}}{\pa t}=\left[-\nabla^{2}+\frac{g_{22}}{g_{11}}\left|\psi_{2}\right|^{2}+\frac{g_{12}}{g_{11}}\left|\psi_{1}\right|^{2}\right]\psi_{2}.
      \label{cgpe3}
    \end{split}
  \end{equation} 
 From now on we work exclusively with dimensionless quantities.  
 
The number of particles in each species and the total energy of the system are conserved within the coupled GPEs.   The number of particles in each species, $N_{i}$, can be expressed in momentum space as
    \begin{eqnarray}
    N_i=V\int n_{\mathbf{k}i}(t)d\mathbf{k},
    \end{eqnarray}
    where $V$ is the volume of the box and $n_{\mathbf{k}i}$ denotes mode occupation numbers of species $i$ (with $n_{0i}$ being the corresponding condensate population).  The system energy is given by the integral
  \begin{equation}
    \begin{split}
      E=\int  & \left(\left|\nabla\psi_{1}\right|^{2}+\frac{1}{2}\left|\psi_{1}\right|^{4}+\left|\nabla\psi_{2}\right|^{2}+\frac{g_{22}}{2g_{11}}\left|\psi_{2}\right|^{4} \right. \\ 
        &\left. +\frac{g_{12}}{g_{11}}\left|\psi_{1}\right|^{2}\left|\psi_{2}\right|^{2}\right)\total\mathbf{r}.
      \label{eq:energy}
    \end{split}
  \end{equation}

For simplicity we consider the gases to exist within a three-dimensional periodic box; then the true condensate of each species is  identified as the zero momentum mode.  The dimensionless coupled GPEs are solved numerically on a cubic grid of size $64^3$ using a fourth-order Runge-Kutta method.  The time step is $\Delta t=0.01$ and the (isotropic) grid spacing is $\Delta = 1$.  This spatial discretization implies that momentum is discretized into discrete modes and that high momenta are not described.  To formalize the latter effect, an ultraviolet cutoff is introduced such that $n_{\mathbf{k}}(t)=0$ for $k> 2\sqrt{3} \pi / \Delta$, where $k=|{\bf k}|$.  This cutoff, which arises numerically in practice, is also required formally to prevent an ultraviolet catastrophe in the classical field model. As long as the grid size exceeds $16^3$, the equilibrium properties predicted by classical field simulations, e.g. final condensate fraction, are effectively independent of grid size~\cite{PhysRevLett.95.263901} (although the time scale to reach equilibrium does depend on grid size).

    \subsection{Equilibrated Single Species Bose Gas}

Our physical scenario begins with a single-species Bose-condensed gas which is already at equilibrium.  Taking this to be species 1 (and dropping the 1 subscript for brevity), we form this equilibrated state via classical field propagation of the GPE for species 1, i.e. Eq. (\ref{cgpe1}) with $\psi_2=0$, from the nonequilibrium initial conditions
  \begin{equation}
    \psi \left(\mathbf{r},t=0\right)=\sum_{\mathbf{k}}a_{k}\exp(i\mathbf{k}\cdot\mathbf{r})
    \label{eq:rand2}
  \end{equation}
where the magnitudes of $a_{k}$ are uniform and the phases are distributed randomly~\cite{PhysRevA.66.013603}.     We characterize the initial state in terms of its particle density $N/V$ and average energy density $\langle E \rangle/V$.  The importance of these quantities is that they determine the equilibrium state, including temperature and condensate fraction, of the single species gas~\cite{PhysRevLett.95.263901}.

\begin{figure}
        \begin{overpic}[scale=0.33]{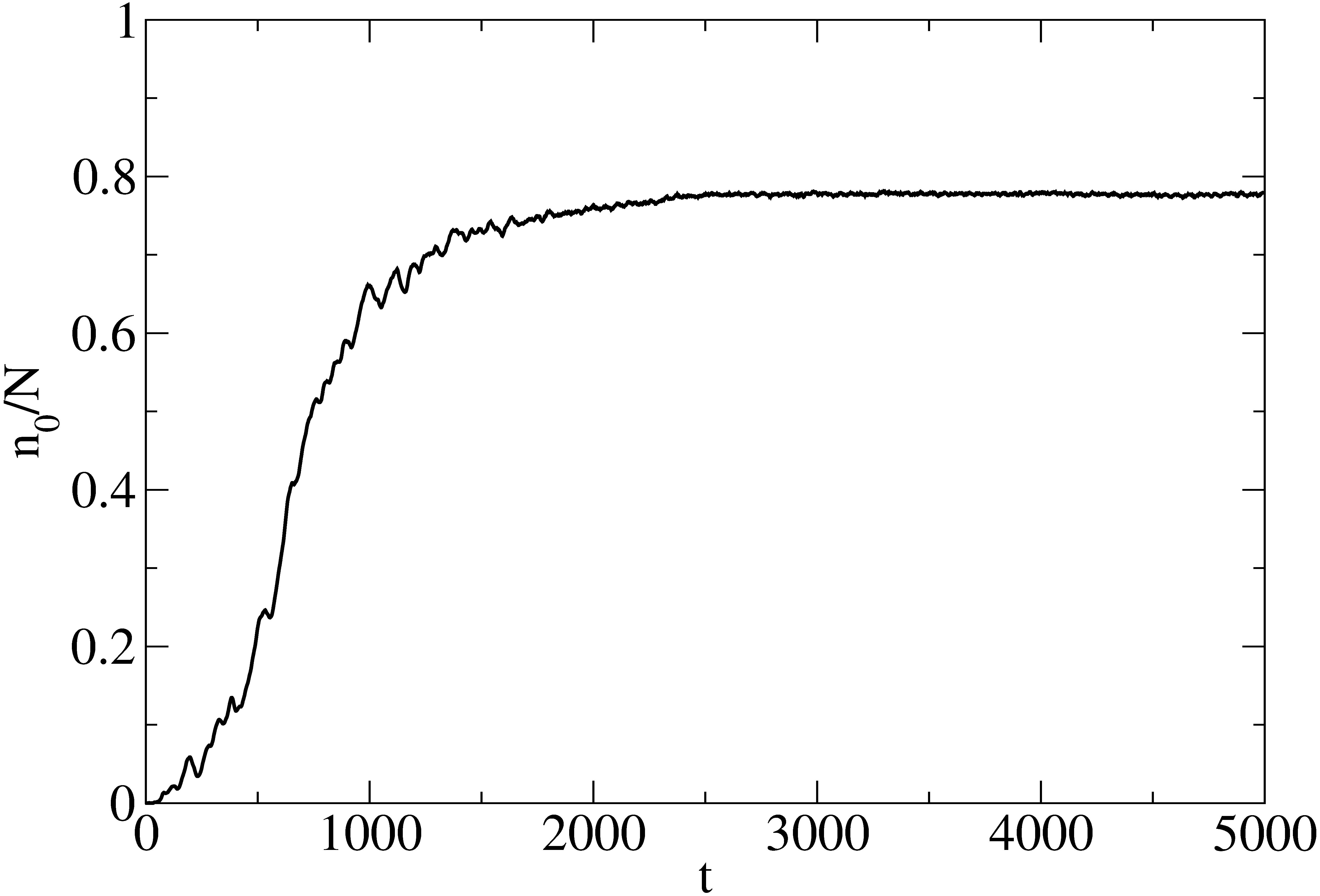} 
         \put(40,10){\includegraphics[scale=0.3]{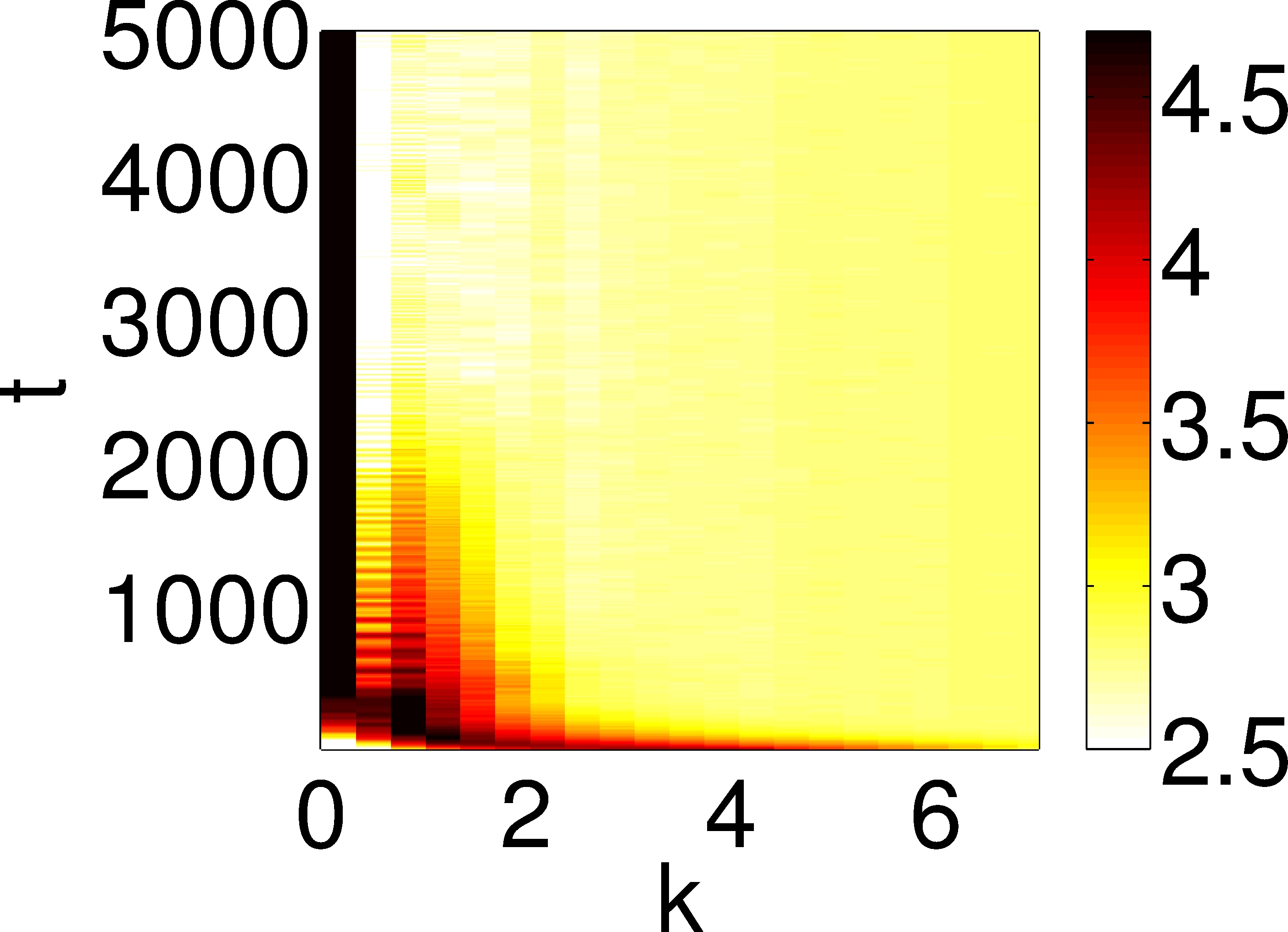}}  
  \end{overpic}
        \caption{(Color online) Evolution of condensate fraction for a homogeneous single-species Bose gas starting from highly nonequilibrium initial conditions.  Parameters:  $N/V=0.5$ and $\langle E\rangle/V=1.2$. Inset: Evolution of the mode occupation $n_{k}(t)$ (plotted as $\log_{10}(n_{k})$).\label{fig:single_frac}}
      \end{figure}

A typical evolution of this single-species system is shown in Fig. \ref{fig:single_frac}, for the arbitrary parameters $N/V=0.5$ and $\langle E\rangle/V=1.2$. Initially the condensate fraction $n_0/N$ is approximately zero since the occupation numbers are distributed across $k$ (see inset).  Rapidly, the occupation numbers condense towards low $k$, resulting in a rapid growth of $n_0/N$.  Later $n_0/N$ approaches a constant value (of $n_0/N = 0.77$ for the arbitrary parameters used here) associated with the equilibrium of the gas.  The final state has macroscopic occupation of the $k=0$ mode, confirming the formation of a condensate (see inset).  

At equilibrium, the quantity $\sum_{k' \leq k} n_{\bf k'}$ shows a characteristic bimodal shape as a function of $k$, as discussed in \cite{PhysRevA.66.013603}.  The lower-$k$ regime is associated with the quasi-condensate and the upper-$k$ regime with the thermal non-condensate.  During the evolution, we observe this bimodal distribution to rapidly emerge, and for the parameters used here we identify the crossover to occur at $k \approx 2.7$.  The norm (particle number) and total energy are not conserved during these dynamics, since energetic particles can ``evaporate" from the system due to the momentum cutoff.  For example, during these dynamics, the total energy decreases by approximately $30\%$ and the norm by $0.1\%$. Consistent with~\cite{PhysRevA.66.013603}, we observe the formation of a turbulent tangle of vortices in the quasi-condensate during thermalization, which decays over time.  

For a different initial energy and particle density the equilibration dynamics show the same qualitative form but approach a different equilibrium state (with different condensate fraction).  For fixed particle density $N/V$, the final condensate fraction decreases as the energy density $\langle E\rangle/V$ of the system is increased~\cite{PhysRevLett.95.263901}.   

At a time $t_{\rm eq}=5000$, when the system is deemed equilibrated, we define the equilibrium state of the single-species gas $\psi_{{1, {\rm eq}}}=\psi_{1}(t=t_{\rm eq})$.  Note that at this time all vortices have decayed from the system.  Our subsequent results are not sensitive to the choice of $t_{\rm eq}$, providing the state is indeed equilibrated.   Note that, unless described otherwise, we use the single-species system employed in Fig. \ref{fig:single_frac} ($N/V=0.5$ and $\langle E\rangle/V=1.2$).

  \subsection{Non-Equilibrium Binary Bose Gas}

Starting from the equilibrated single-species gas at time $t_{\rm eq}$, we instantaneously transfer a proportion $\alpha$ ($0\leq\alpha\leq 1$) of the particles into species 2 according to the transformations
  \begin{equation}
    \begin{split}
      &\psi_{1}({\bf r}, t=t_{\rm eq})=\sqrt{1-\alpha}\psi_{\rm eq}({\bf r})+\eta_1({\bf r}),\\
      &\psi_{2}({\bf r},t=t_{\rm eq})=\sqrt{\alpha}\psi_{\rm eq}({\bf r})+\eta_2({\bf r}).
    \end{split}
\label{eqn:split}
  \end{equation}
Low level noise, $\eta_1({\bf r})$ and $\eta_2({\bf r})$, uniformly randomly distributed about zero with maximum amplitude $5\times10^{-4}$ (where the mean amplitude of $\psi_i$ is $1$ in our dimensionless units) is applied to each species to break symmetries. Our qualitative findings are independent of the noise amplitude.     
The subsequent dynamics of the two coupled gases are obtained via propagation of the coupled GPEs~\eqref{cgpe3}.   Note that, immediately after splitting, the total kinetic energy density is essentially unchanged (to within the effects of the noise terms $\eta_{i}$), that is, each species is formed with the same average kinetic energy per particle as in the original single-species condensate.  The interaction energies following the population transfer depend on the interaction parameters.  

Following the work of Berloff for miscible homogeneous binary Bose gases \cite{Berloff_2006,Salman20091482}, in our work we also identify  the condensate fraction in each species as the population of the $k=0$ mode, having chosen to extend this identification to the immiscible regime as well.  This regime is characterized by formation of phase-separated domains in which one species effectively confines the other.  Thus, while each species may be locally homogeneous within each domain, it is inhomogeneous on the larger scale.  Previous work~\cite{Astrakharchik,Muller} on single component Bose gases which are inhomogeneous due to external trapping has revealed the importance of condensate deformation and the deviation of the Penrose-Onsager condensate mode \cite{penrose,Carsten,Blakie,Wright2011} (obtained by numerical diagonalization) from the $k=0$ mode.  In the phase-separated binary Bose gas, we can expect the inhomogeneous density to have a qualitatively similar effect on the nature of the condensate fraction.  Nonetheless we expect that the $k=0$ mode will provide an important and insightful metric to characterise the thermodynamic state of the gas.

  \section{Equilibration Properties of Binary Bose Gases \label{sec:results}}

We seek to establish how the binary Bose gas, formed instantaneously via population transfer, evolves from its non-equilibrium state and what the  equilibrium properties are.  The key parameters that will dictate the thermodynamical evolution are the population transfer $\alpha$, the intraspecies interactions and the interspecies interactions.   Below we explore the dependencies on these parameters.

    \subsection{Equal Intraspecies Interactions and Equal Populations}
We begin by considering the intraspecies interactions to be equal $g_{11}=g_{22}$ and for each species to have the same population ($\alpha=0.5$).    We can write the immiscibility criteria (\ref{eqn:crit}) as $g_{12}/g_{11}>\sqrt{g_{22}/g_{11}}$ (assuming repulsive intraspecies interactions), which simplifies to $g_{12}/g_{11}>1$ for the intraspecies interactions considered here.  

\begin{figure}[!h]
        \subfigure{\includegraphics[width=0.48\textwidth,clip]{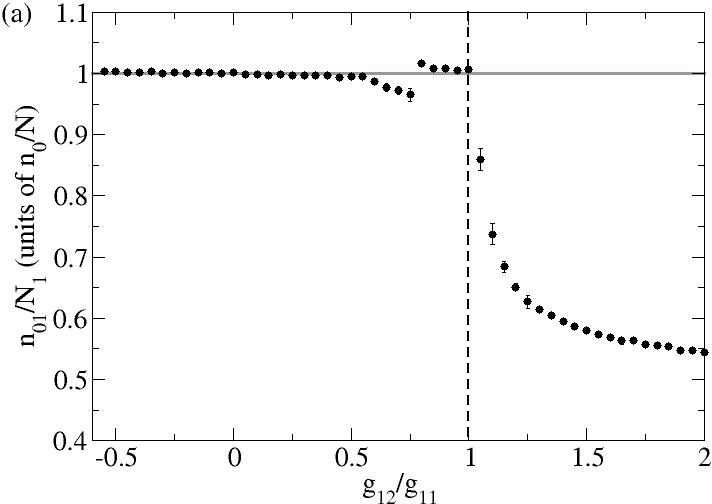}}
        \subfigure{\begin{overpic}[scale=0.35]{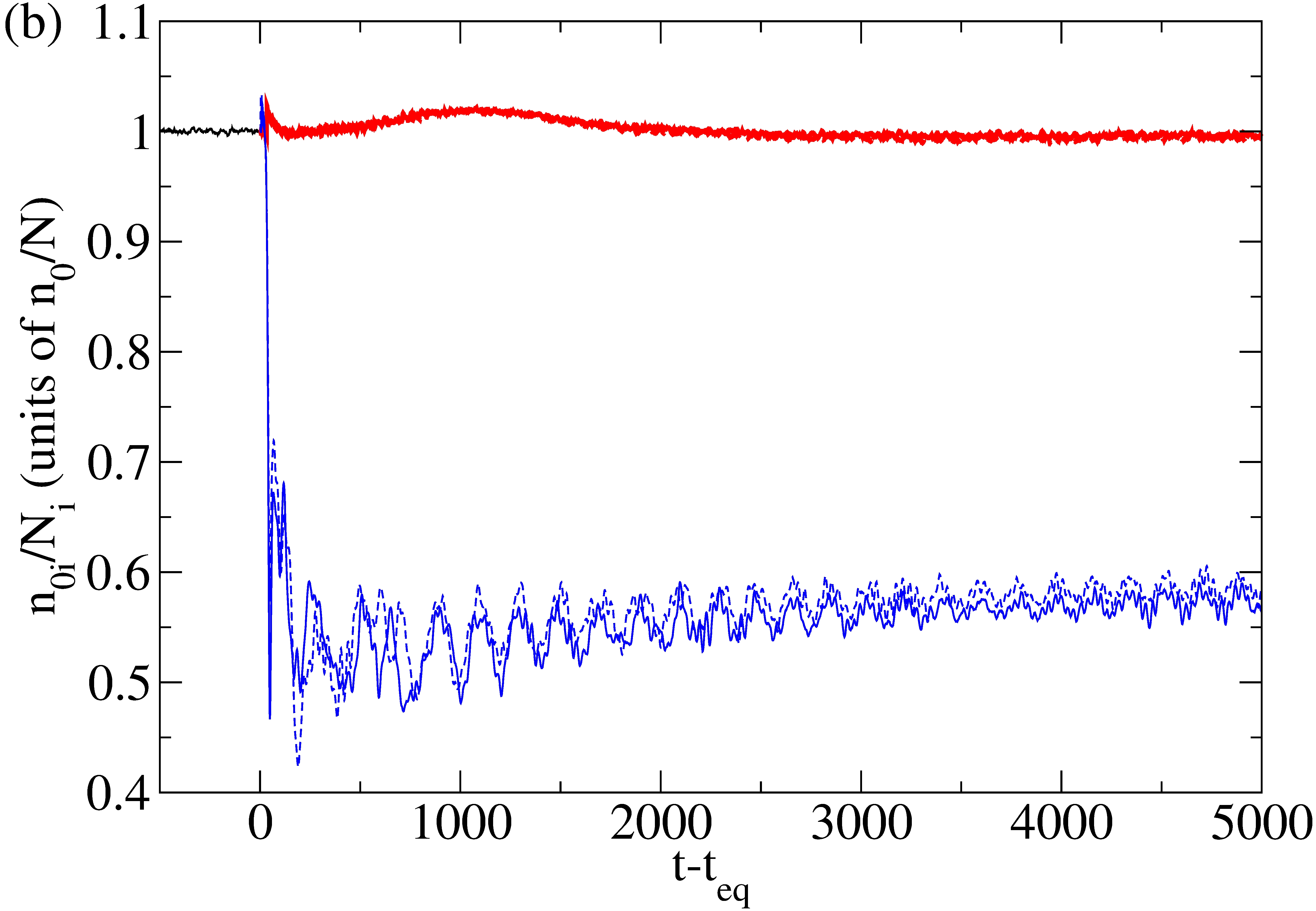} 
         \put(21,26){\includegraphics[scale=0.09]{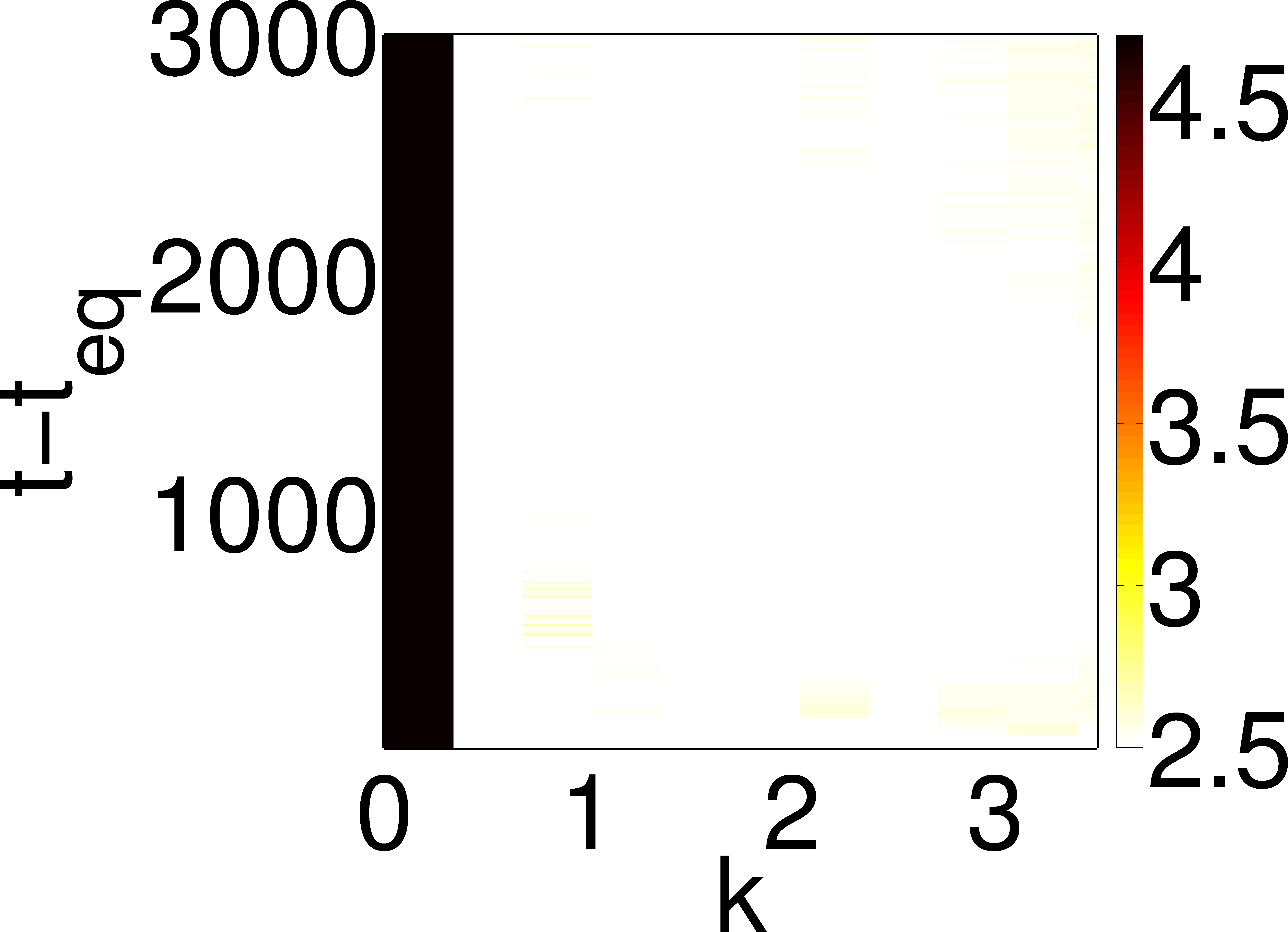}}
         \put(58,26){\includegraphics[scale=0.09]{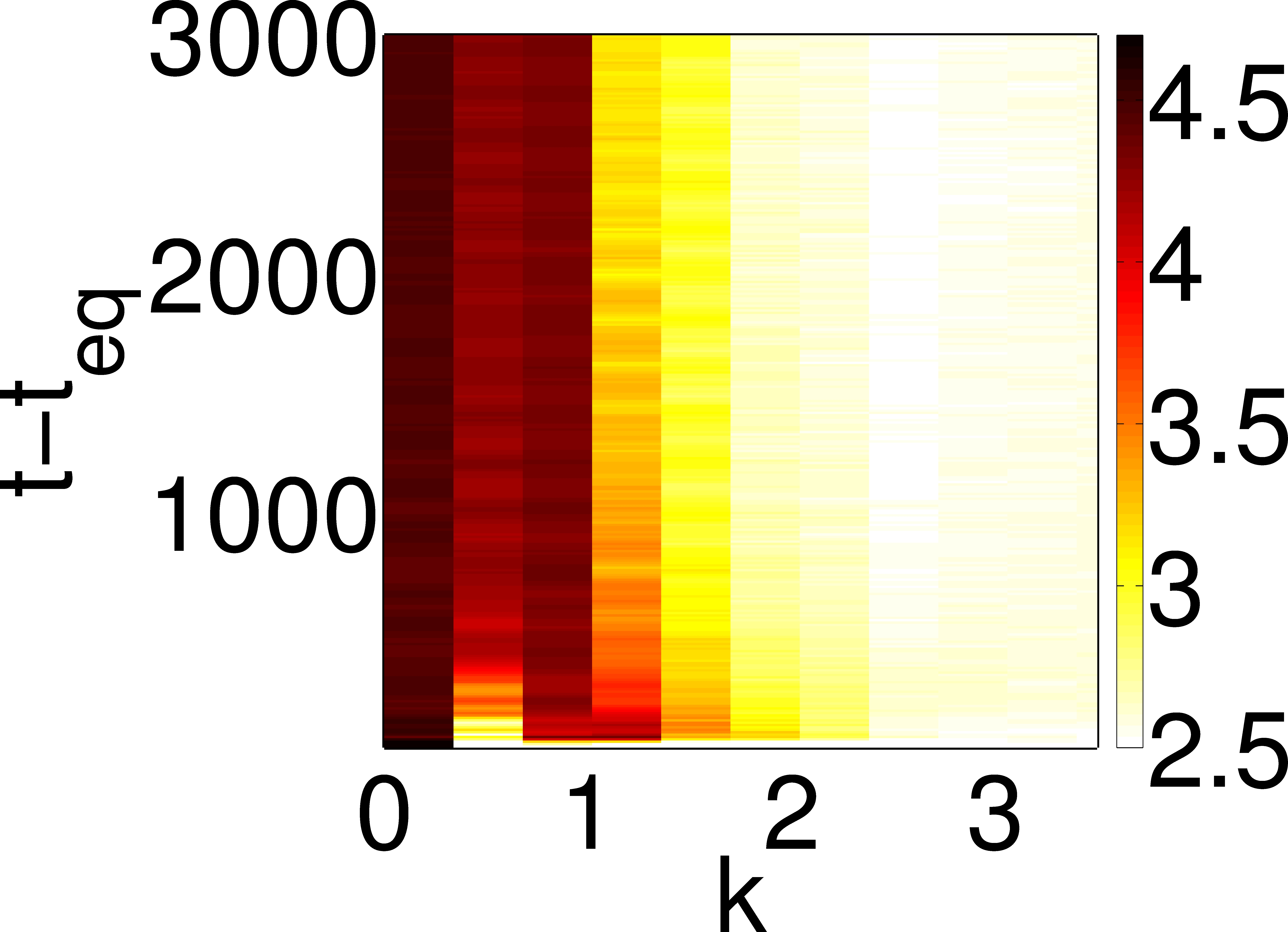}} 
        \end{overpic}}
    \caption{(Color online) Equilibration of a single-species homogeneous Bose gas which is instantaneously transferred (at $t=t_{\rm eq}$) into a binary Bose gas with equally populated components ($\alpha=0.5$) and equal intraspecies interactions ($g_{11}=g_{22}$).  (a) Final equilibrium condensate fraction of species 1 as a function of the interspecies interactions $g_{12}/g_{11}$.  Condensate fraction is expressed in units of $n_0/N$, the equilibrated condensate fraction of the single-species gas.  Error bars give standard deviation over $1000$ time units, with most being obscured by the markers. The data for species 2 is identical (to within statistical uncertainty) of that of species 1. The initial single species condensate fraction is shown by the horizontal grey line; the miscible-immiscible threshold by the vertical dashed line.
(b) Evolution of condensate fraction for a miscible case ($g_{12}=0.5$; thick red curves, top) and an immiscible case ($g_{12}=1.5$; thin blue curves, bottom).  Solid (dashed) curves represent species 1 (2). The  initial single-species condensate fraction is shown for $t<t_{\rm eq}$ (black curve). Insets: Corresponding evolution of the mode occupation for species 1 (plotted as $\log_{10}(n_{k1})$) for the miscible (left) and immiscible (right) cases.\label{fig:equal}}
      \end{figure}

Our general findings are summarized in Fig.~\ref{fig:equal}(a), which plots the equilibrium condensate fractions $n_{0i}/N_{i}$ for each species as a function of the interspecies interactions $g_{12}$ (expressed in units of $g_{11}$).  The condensate fractions are scaled by their initial values, that is, the condensate fraction of the equilibrated single-species gas ($n_0/N=0.77$ for the arbitrary parameters we employ here).  The final equilibrium condensate fractions, according to the ergodic theorem, are determined by time-averaging the condensate fractions (here over 1000 time units) following their saturation to a constant average value.  Note that the size of the thermal fluctuations, characterized by the standard deviation of this data, are small.  Note also that the results for species 1 and species 2 are indistinguishable to within the statistical noise (see Fig.~\ref{fig:equal}(b)) and so only the results for species 1 are presented in  Fig.~\ref{fig:equal}(a).

It is evident from Fig. ~\ref{fig:equal}(a) that the equilibrium state of the system depends critically on whether the two-species system is miscible ($g_{12}/g_{11}\leq 1$) or immiscible ($g_{12}/g_{11} > 1$).  We now discuss each case in turn.

\subsubsection{Miscible regime}
For $g_{12}/g_{11}\leq 1$ the final condensate fraction in each of the two condensates remains close to that of the original condensate (represented by the horizontal grey line).  To illustrate the dynamics in more detail, we show how the condensate fraction evolves in Fig.~\ref{fig:equal}(b) for the example value $g_{12}/g_{11}=0.5$ (thick red lines, top). Immediately following the population transfer at $t=t_{\rm eq}$, there is a sharp spike in the condensate fraction; this feature, which we observe for all simulations, is associated with the small but noticeable energy driven into the system by the noise terms $\eta_i$ in the transformations (\ref{eqn:split}).  We have confirmed that reducing the noise amplitude reduces the amplitude of this feature but does not affect the equilibrium properties.  There is a subsequent small modulation of the condensate fractions, but these settle to values which are very close to the original.  During the evolution the mode occupations (left inset) undergo no visible changes, being dominated throughout by the macroscopic population of the $k=0$ mode.  

We observe similar dynamics for all miscible cases ($g_{12}/g_{11}\leq 1$), with the final equilibrium condensate fraction differing by at most a few percent from the original condensate.  In general, the population transfer process modifies the total energy per particle: while the kinetic energy per particle is unchanged, the self-interaction energy becomes reduced (due to the quadratic scaling of self-interaction energy with particle density) and interspecies interaction energy is introduced.  Over time there is then a weak equilibration within each species.  

Within the miscible regime there exist two special cases.  For $g_{12}/g_{11}=0$ the two species evolve independently over time.   For $g_{12}/g_{11}=1$, and assuming $g_{11}=g_{22}$, the energy-per-particle in each species is the same as in the original Bose gas (and with the same proportion of kinetic and interaction energy), barring the weak effect of noise introduced during splitting.  For both of these special cases the final condensate fraction is identical (within noise) to that of the original single-species Bose gas.    

\subsubsection{Immiscible regime}
For $g_{12}/g_{11}>1$ the equilibrium condensate fractions are dramatically reduced (Fig.~\ref{fig:equal}(a)).  Consider the example case for $g_{12}/g_{11}=1.5$ (thin blue lines in Fig.~\ref{fig:equal}(b), bottom).  Following the population transfer, the condensate fractions drop rapidly (over tens of time units), levelling out at around $58\%$ of their initial value.   The origin of this reduction is as follows.  The population transfer forms two immiscible species with overlapping density profiles (identical apart from the imposed noise) and drives interaction energy into the system.  From this energetically unfavourable state, the strong and repulsive interspecies interactions induce the species to phase-separate.  This process imparts opposing momentum to the species, which is evident in the sudden population of finite-$k$ modes following population transfer (Fig.~\ref{fig:equal}(b), right inset).  In doing so, interspecies interaction energy becomes converted into kinetic energy, raising the average kinetic energy per particle, and leading to a lower final condensate fraction.  For the miscible case, the population transfer can also drive interaction energy into the system but, crucially, there is no subsequent phase separation and hence no significant growth of kinetic energy. As discussed earlier, in the immiscible regime, the deformation of a species due to the effective confining by the other may lead to the population of the $k=0$ mode not coinciding with the Penrose--Onsager condensate. 

The results presented in Fig.~\ref{fig:equal}(a) are independent of the original single-species condensate fraction (hence the motivation for scaling the $y$-axis in terms of this quantity).  We have verified this by starting from single-species condensates with different condensate fractions (obtained  from starting the single-species simulation with different energy densities), and find the results to be consistent with the data in Fig.~\ref{fig:equal}(a).   In other words, the final condensate fraction is a fixed ratio of the initial fraction, depending only on the interactions $g_{12}/g_{11}$.

    \subsection{Equal Intraspecies Interactions but Unequal Populations}

\begin{figure}[!h]
        \subfigure{\includegraphics[width=0.48\textwidth,clip]{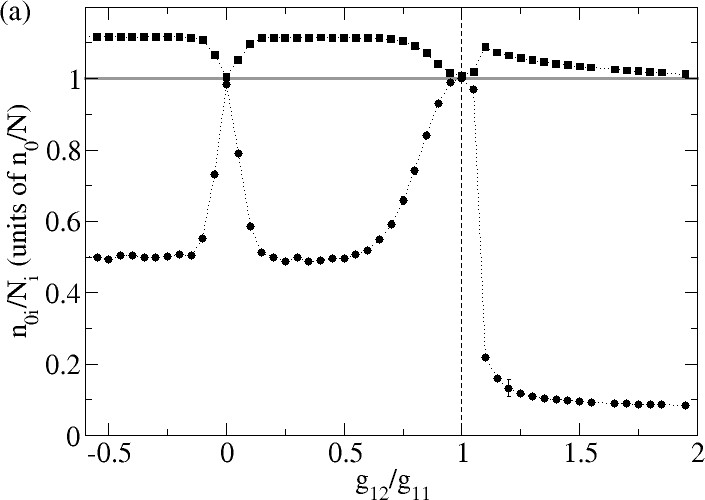}}
         \subfigure{\includegraphics[width=0.48\textwidth,clip]{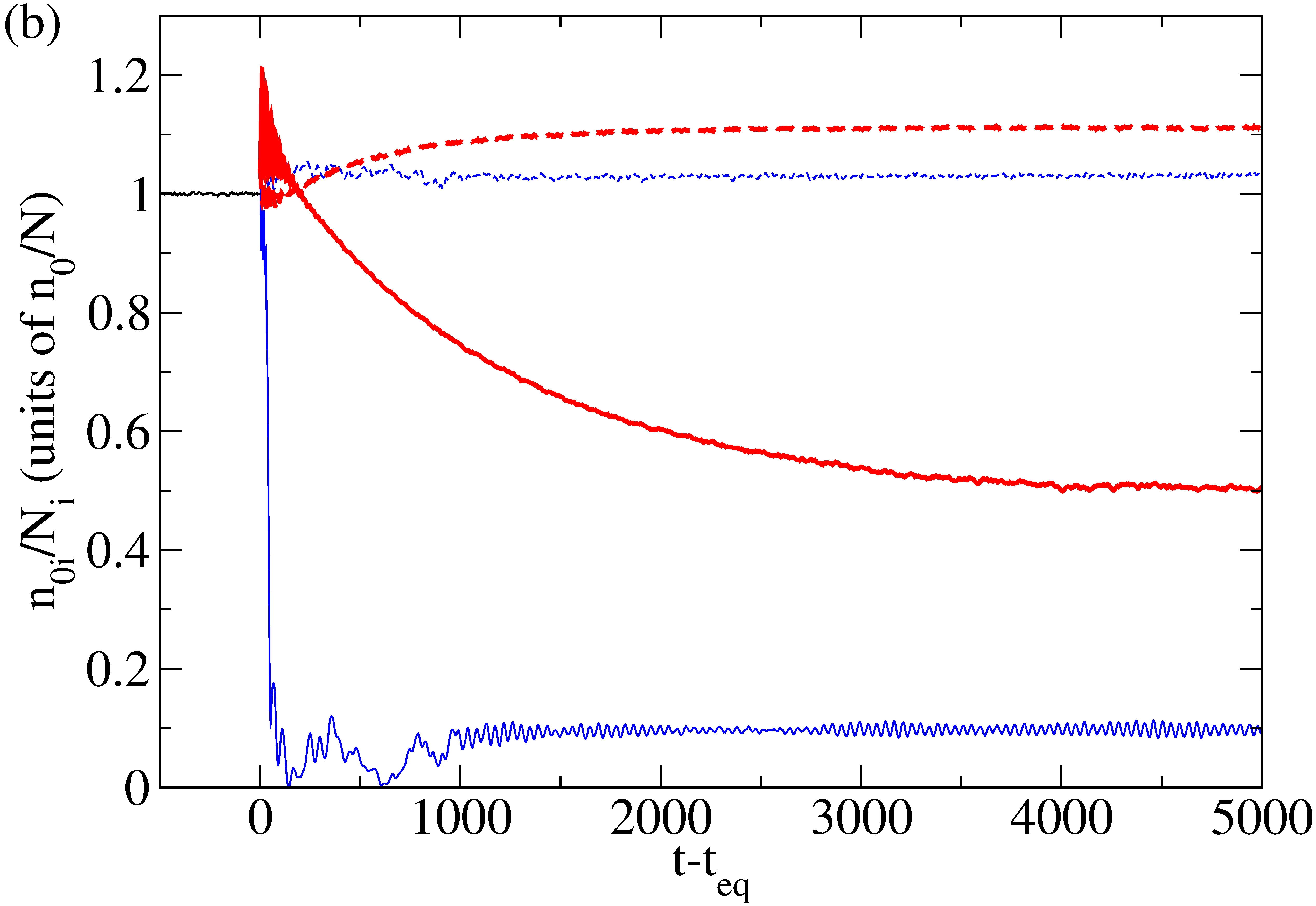}}
        \caption{(Color online) Equilibration of a single-species homogeneous Bose gas which is instantaneously transferred (at $t=t_{\rm eq}$) into binary Bose gas with unequal populations ($\alpha=0.8$) and equal intraspecies interactions ($g_{11}=g_{22}$). (a) Final equilibrium condensate fractions for species 1 (circles) and species 2 (squares).  Most error bars are obscured by the markers. (b) Evolution of condensate fraction for a miscible case ($g_{12}=0.5$; thick red curves) and an immiscible case ($g_{12}=1.5$; thin blue curves).  Solid (dashed) curves represent species 1 (2). The  initial single-species condensate fraction is shown for $t<t_{\rm eq}$ (black curve).\label{fig:unequal}}
      \end{figure}

 We now move on to consider unequal populations and consider the example case of $\alpha=0.8$.  The final condensate fraction, shown in Fig.~\ref{fig:unequal}(a), shows a more complicated dependence on $g_{12}/g_{11}$ than for equal populations (Fig.~\ref{fig:equal}(a)).  In particular, we now see that species 1 (circles), which has the smaller number of particles, often has a lower condensate fraction than species 2 (triangles).   What's more, species 2 can develop a condensate fraction which is slightly {\it higher} than the original condensate (grey line).  The special nature of the values $g_{12}/g_{11}=0$ and $1$ is even more evident for unequal populations; only here do the condensate fractions become equal to each other and equal to the original condensate.

In Fig.~\ref{fig:unequal}(b) we show the evolution for a miscible case, $g_{12}/g_{11}=0.5$ (thick red curves).  After the population transfer and following the noise-induced cusp, the condensate fraction $n_{i,0}(t)/N_i$ of species 1 decays slowly to around $50\%$ of its initial value, while that of species 2 increases by approximately $10\%$.  For the immiscible case, $g_{12}/g_{11}=1.5$ (thin blue curve), we see the much faster and larger decay of species 1 condensate fraction, reducing by around $90\%$, while that of species 2 increases slightly (a few percent).  These curves are typical of the behaviour in the miscible and immiscible regimes, respectively, providing one is away from the special values of $g_{12}$.  

In the immisicible regime, the condensate in species 1 mostly ``boils off'', leaving behind only a weakly populated condensate.  This is again associated with the phase separation dynamics which proceed the population transfer.  Here, for unequal splitting, species 1 gains more kinetic energy per particle than species 2, due to the low total mass/inertia of species 1.

    \subsection{Equal Intraspecies Interactions and General Splitting}

\begin{figure}[!h]
        \includegraphics[width=0.48\textwidth,clip]{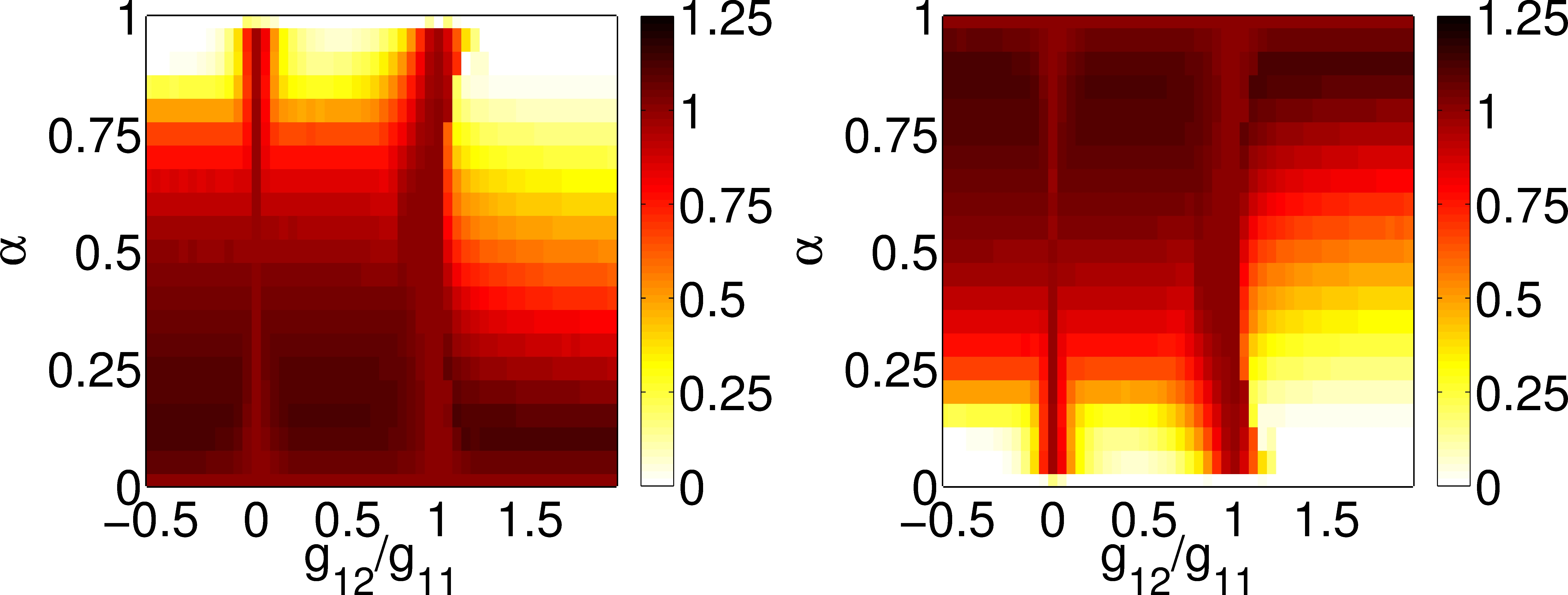}
        \caption{(Color online) Equilibrated condensate fractions (units of $n_{0}/N$) for species 1 (left) and species 2 (right) as a function of the population transfer $\alpha$ and the interspecies interactions $g_{12}/g_{11}$. The species are assumed to have equal intraspecies interactions ($g_{11}=g_{22}$).\label{fig:frac_equal}}
      \end{figure}

Figure~\ref{fig:frac_equal} shows the equilibrium condensate fractions (scaled by their initial value) in each species as a function of both the population transfer $\alpha$ and the interspecies interactions $g_{12}$.  Light/dark areas represent low/high condensate fractions.  The two special values of interspecies interactions, $g_{12}/g_{11}=0$ and $1$, are even more apparent since the condensate fractions here remain unchanged for all values of $\alpha$.  The values $\alpha=0$ and $\alpha=1$ represent special cases of the population transfer for which only one species is relevant, while $\alpha=0.5$ (discussed earlier) is special in that both species have identical dynamics and properties.  In the rest of the parameter space the general trend as $\alpha$ increases is for a reduction in the condensate fraction of species 1 and an increase of it in species 2.  For sufficiently extreme population transfer ($\alpha \gtsimeq 0.8$ or $\alpha \ltsimeq 0.2$) the condensate fraction of the smaller component diminishes to around zero (white regions in Fig.~\ref{fig:frac_equal}) and the condensate fraction in the larger component can slightly exceed the original value. In the immiscible regime ($g_{12}/g_{11}>1$) this extends to more moderate values of $\alpha$.

    \subsection{Unequal intraspecies interactions}

All of the results discussed so far have been for equal intraspecies interactions, $g_{11}=g_{22}$.  For $g_{11} \neq g_{22}$ the results are qualitatively similar, except with the miscible and immiscible regimes becoming shifted according to the immiscibility criterion $g_{12}/g_{11}>\sqrt{g_{22}/g_{11}}$.  For the $|F=1,m_F=-1\rangle$ and $|F=2,m_F=1\rangle$ hyperfine states of $^{87}$Rb, $g_{11}$ and $g_{22}$ differ by less than $5\%$ \cite{Egorov} such that this shift is small.

  \section{Conclusions\label{sec:conc}}

Starting from an equilibrated Bose-condensed gas, we have considered the experimentally-motivated scenario in which a proportion of the atoms are coherently transferred into a different species, thereby forming a non-equilibrium binary Bose gas mixture.  Using classical field simulations in a homogeneous box we have analyzed the ensuing dynamics and final state of each species, highlighting the strong dependence on whether the components are miscible or immiscible.  For equally populated components, the final condensate fractions vary little from that of the original Bose gas when the components are miscible, but reduce significantly when the components are immiscible.  This can be related to the phase separation dynamics which follow the population transfer and which cause a growth in kinetic energy at the expense of the interspecies interactions.  For unequal populations the general trend is for the smaller component to undergo a reduction in condensate fraction, which for even moderate parameters can lead to the condensate fraction becoming effectively zero; meanwhile the condensate fraction of the larger components typically increases, albeit by a smaller amount.  

In considering a homogeneous system we have revealed the rudimentary role of interactions on the equilibration dynamics.  In experiments, however, Bose gases are confined in trapping potentials, usually harmonic in shape, which further influence the system through the imposition of an inhomogeneous density profile and a boundary.  While this significantly changes the stationary density distributions of binary condensates, e.g., leading to ball-in-shell phase-separated states \cite{PhysRevLett.77.3276,Pu,PhysRevLett.81.1539,Trippenbach,Gautum,Pattinson2013}, and their dynamical properties, e.g. quench dynamics \cite{Lui} and shape oscillations  \cite{PhysRevLett.81.243,PhysRevLett.81.1539}, the atomic interactions remain a dominant influence on the system.   Importantly, the miscible and immiscible regimes, which drive our main observations in the condensate fractions, persist in the presence of a trap (albeit with the crossover shifted by the effects of quantum pressure arising from the inhomogeneous density \cite{Wen}). Indeed, for the scenario of a single-species Bose gas the equilibration dynamics have been shown to be qualitatively similar between homogeneous \cite{Davis,Davis2} and trapped systems \cite{Blakie2005}.    As such, we can expect our results to also provide {\em qualitative} insight into the equilibration of trapped Bose gases.  Our chosen identification of the condensate as the $k=0$ mode (which does not account for condensate deformation induced by phase-separated density inhomogeneities), could be  improved by defining the condensate as the macroscopically-occupied eigenmode (which in practice requires numerically-intensive diagonalization of the one-body density matrix).  This would also be crucial for considering the experimentally relevant scenario of harmonic external confinement of the gas.  In this context we remark that recent studies of quenched immiscible trapped binary condensates indicate that the condensate fractions are practically independent of the actual spatial distribution of the condensates \cite{Lui}. 



The specific scenario we consider of a binary Bose gas with equal masses has been performed experimentally with trapped Bose gases of $^{87}$Rb, most commonly utilizing the hyperfine states $|F=1,m_F=-1\rangle$ and $|F=2,m_F=1\rangle$ \cite{PhysRevLett.81.1539,PhysRevLett.81.1543,PhysRevLett.99.190402,McGuirk,PhysRevA.80.023603}.   This system has $g_{12}^2=1.002 g_{11}g_{22}$ \cite{Egorov} and lies just in the immiscible regime.  Our results suggest a sizeable reduction in condensate fraction of at least one of the two components, intrinsic to the rapid population transfer procedure performed in experiments.  In future work, we hope to extend our analysis to trapped Bose gases to determine more precisely the role of interactions and trapping on the equilibration of the system.

  \begin{acknowledgements}
    This work made use of the facilities of N8 HPC provided and funded by the N8 consortium and EPSRC (Grant No.EP/K000225/1). The Centre is coordinated by the Universities of Leeds and Manchester. NPP acknowledges funding from EPSRC (Grant No.EP/K03250X/1).
 \end{acknowledgements}


\begin{thebibliography}{10}
\bibitem{spinor_remark} Spinor condensates possess a spin degree of freedom and may be considered as two or three-component condensates but where each species is associated with a different spin state \cite{Ueda}.
\end{thebibliography}
\end{document}